\documentclass[preprintnumbers,superscriptaddress,amsmath,amssymb,prd,nofootinbib,onecolumn]{revtex4-2}
\usepackage{amssymb}
\usepackage{mathrsfs}
\usepackage{amsfonts}
\usepackage{amsthm}
\usepackage{braket}
\usepackage{bm}
\usepackage{qcircuit}
\usepackage{graphicx}
\usepackage{amsfonts}
\usepackage[colorlinks,linkcolor=red,anchorcolor=blue,citecolor=green]{hyperref}
\usepackage{subfigure}
\usepackage{float}
\usepackage{enumerate}
\usepackage{multirow}

\newcommand\e{\mathrm{e}}
\usepackage{array}
\usepackage[figuresright]{rotating}
\usepackage{upgreek}

\begin{document}
	
	\title{Thermodynamics of the \texorpdfstring{\(H^{2}+H^{-2}\)}{H\textasciicircum 2 + H\textasciicircum -2} Dark Energy model}
	
	\author{Guo Chen}
	\email{petterchenguo@qq.com}
	\affiliation{The Shanghai Key Lab for Astrophysics, 100 Guilin Rd, Shanghai 200234, P.R.China}
	\affiliation{Department of Physics, Shanghai Normal University,
		100 Guilin Rd, Shanghai 200234, P.R.China}
	
	\author{Chao-Jun Feng}
	\thanks{Co-first author}
	\email{fengcj@shnu.edu.cn}
	\affiliation{Department of Physics, Shanghai Normal University,
		100 Guilin Rd, Shanghai 200234, P.R.China}
	
	\author{Wei Fang}
	\thanks{Corresponding author}
	\email{wfang@shnu.edu.cn}
	\affiliation{The Shanghai Key Lab for Astrophysics, 100 Guilin Rd, Shanghai 200234, P.R.China}
	\affiliation{Department of Physics, Shanghai Normal University,
		100 Guilin Rd, Shanghai 200234, P.R.China}

	\author{Chenggang Shu}
	\affiliation{The Shanghai Key Lab for Astrophysics, 100 Guilin Rd, Shanghai 200234, P.R.China}

	\begin{abstract}
		In this study, we introduced a phenomenological dark energy model (\(H^{2}+H^{-2}\)  model) that incorporates the first-order approximation of Kaniadakis holographic entropy dark energy and utilizes the Hubble horizon, \(1/H\) as the infrared cutoff. The advantage of this model is that it can relieve the Hubble tension issue and cicumventing the potential issue of a "big rip". In this article, we will study the thermodynamics of the model and obtain the corrected temperature. In addition, we found that the model modified the entropy area relationship by adding an area cubic term in addition to the area term.
	\end{abstract}

	\maketitle

	\section{INTRODUCTION}
	Nowadays, there is increasing evidence suggests that the universe is undergoing accelerated expansion. To explain this phenomenon, many models have been proposed from two directions. One is to modify General Relativity(GR) and change the expression of gravity; Another approach is to add a negative pressure substance called dark energy(DE) to the composition of cosmic matter. The simplest model among them is the cosmological constant (\( \Lambda\)CDM) model. In this model, the energy density of dark energy is constant, and the state equation parameter \(\omega =-1\). However, this model has an unavoidable fine-tuning issue, and recent observational evidence suggests that the parameters of the dark energy state equation may not be constant. Therefore, the \(\Lambda\)CDM model cannot well conform to the evolutionary phenomena of the universe.
	
	Recently, holographic dark energy models have become a popular alternative method for solving the problem of accelerated expansion \cite{Li_2004,Wang_2017}. This model is based on the holographic principle, which asserts that the entropy of the universe should not exceed the entropy of black hole of the same size. Therefore, the energy density, which is inversely proportional to the square of the infrared cutoff \(\sim 1/L^2\)), is constrained. Recently, This principle has been explored in the context of thermodynamics \cite{Manoharan2023}. In addition, in Ref.\cite{Zhang2022,Sharma_2022}, the R\'enyi holographic dark energy model was discussed, which examines the interaction between dark energy and dark matter in fractal cosmology. In Ref.\cite{Feng:2009jr} , the author also introduced viscous fluids, significantly alleviating the issue of the Universe's age. In Ref.\cite{Feng:2008hk}, the author reconstructed the \(f(R)\) theory model based on the holographic DE. The latest progress in holographic DE models is presented in Ref.\cite{Wang:2016och} and information on DE and modified gravity is also provided in Ref.\cite{Bahamonde:2017ize}.
	
	Kaniadakis intruduced a single-parameter generalization of the Boltzmann-Gibbs entropy, called the Kaniadakis entropy \cite{Kaniadakis_2002,Kaniadakis_2005}, which is defined as\cite{Abreu_2016,Abreu_2017,Abreu_2018,Abreu_2019,Yang_2020,Abreu_2021}:

	\begin{eqnarray}
	S_{k}=-k_{B}\sum_{i}n_{i}\ln_{{k}}{n_{i}}\label{1.1} \,,
	\quad \text{or} \quad S_{k}=-k_{B}\sum_{i}^{W}\frac{P_{i}^{1+K}-P_{i}^{1-K}}{2K} \,,
	\end{eqnarray}
	
	where \(P_{i}\) represents the probability of a specific microstate of the system, and W represents the total number of possible configurations. When applied within the context of black hole physics, the entropy takes the form\cite{Drepanou_2022}:
	
	\begin{eqnarray}
	S_{k}=\frac{1}{K}\sinh (KS_{BH}) \label{1.2} \,,
	\end{eqnarray}
	
	which, in the approximation where \(K\ll 1\), can be simplified to\cite{Drepanou_2022}:
	
	\begin{eqnarray}
	S_{K}=S_{BH}+\frac{K^{2}}{6}S_{BH}^{3}+ \mathcal{O}(K^4)\label{1.3}\,,
	\end{eqnarray}
	
	According to the holographic principle, we can derive the so-called Kaniadakis holographic dark energy (KHDE), whose energy density is expressed as\cite{Fang_2024}:
	
	\begin{eqnarray}
	\rho_{de} = 3(\alpha L^{-2} +\tilde{\beta}  L^2)\,,
	\label{1.4}
	\end{eqnarray}
	
	where \(L\) serves as an infrared cutoff. In many recent literatures, the future event horizon was frequently chosen as the cut-off boundary, as shown in Ref. \cite{Almada_2022,Drepanou2022}. Specifically, for the KHDE model, we adopted \(L = R_h = a\int_t^\infty ds/a(s)\) , a method also referenced in Ref. \cite{Ghaffari2022} within the Brans-Dicke framework. In addition, the dynamic characteristics of Kaniadakis holographic dark energy were scrutinized in \cite{Hernández2022}. In another study \cite{Luciano2022}, the authors delved into the applicability of Kaniadakis statistics as a primary paradigm for characterizing complex systems within the realm of relativity. The justification behind selecting the future event horizon is that in the initial holographic DE model, the Hubble horizon was not sufficient to propel the universe's accelerated expansion. However, the incorporation of an additional term in Eq. (\ref{1.3}) modifies this situation.A similar study was conducted in Ref.\cite{Rani_2022}, but it failed to solve the Hubble tension issue.
	
	In the previous research\cite{Fang_2024}, we utilize the Hubble horizon as the infrared cutoff for the KHDE. The energy density is expressed as
	
	\begin{eqnarray} 
	\rho_{de} = 3(\alpha H^{-2} +\tilde{\beta}  H^2)\,,
	\label{1.5}
	\end{eqnarray}
	
	We examined the evolutionary trajectory of the universe and discoverd that our model successfully explained the observed accelerated expansion. After performing the parameter fitting process, we also found that the current Hubble paramter $H_0=72.8$ km/s/Mpc, revealed there is no Hubble tension in this model. The transition from matter dominance to DE dominance occurs approximately at a redshift of $z \approx 0.419$.The estimated age of the universe is $14.2$ Gyr, basing on the best-fit parameter values. In addition, we also conducted statefinder and $Om$ diagnostic analyses to validate and characterize this model. In the future, the energy density of DE will asymptotically approach a constant, and its behavior wii be much similar to that of the cosmological constant, thus, there will be no issue of a cosmic "big rip."
	
	In this paper, we will use the temperature \(T_{t}\) defined on the trapping horizon of the FRW Universe. We study the thermodynamic behavior of our model, and discovered a correction factor for the entropy-area relation, as well as a new area cubic term.
	
	This paper is organized as follows: In Sec. \ref{sec2}, we give a brief review on the unified first law and its application. In Sec. \ref{sec3}, we study the thermodynamics of the \(H^{2}+H^{-2}\) Dark Energy model. In Sec. \ref{sec4}, we use the best fitting values from Ref. \cite{Fang_2024} to study the evolutionary properties of this model. In the last section, we will give some conclusions and discussions.

	\section{BRIEFLY REVIEW ON THE UNIFIED FIRST LAW\label{sec2}}
	
	According to the general definition of the dynamics of black holes on the trapping horizon, Einstein equations can be written in the form of the "Unified First Law", which was proposed by Hayward\cite{Hayward_1938,Hayward_1994,Hayward_1998,Hayward_1999} and developed by Cai et al.\cite{Cai_Rong_Gen_and_Cao_Li_Ming_2007,Rong_Gen_Cai_2007,Rong_Gen_Cai_2008,PhysRevD.78.124012}. In this section, we will outline the unified first law in (3+1)-dimensional spherically symmetric spacetime, where the metric can be locally written in the form of double-null:
	
	\begin{eqnarray} 
	ds^{2}=-2e^{-f}d\xi^{+}d\xi^{-}+r^{2}d\Omega^{2},
	\label{2.1}
	\end{eqnarray}
	
	Here, \(d\Omega ^{2}\) is the line element of the two-sphere with unit radius, \(r\) and \(f\) are functions of \((\xi^{+}, \xi^{-})\). Each symmetric sphere has two preferred normal directions, namely the null directions \(\partial / \partial \xi ^{\pm}\), which will be assumed future-pointing in the following. And also, we will assume the spacetime is time-orientable. The expansion formula for the radial null geodesic congruence is defined by
	
	\begin{eqnarray} 
	\theta_{\pm } =2r^{-1}\partial _{\pm}r,
	\label{2.2}
	\end{eqnarray}
	
	Here, \(\partial _{\pm}\) denotes the coordinate derivative along \(\xi ^{\pm}\). The expansion measures whether the light rays normal to the sphere are diverging \((\theta_{\pm} > 0)\) or converging \((\theta _{\pm} < 0)\), namely, whether the sphere is increasing or decreasing in the null directions. Note that, although the value of \(\theta _{\pm}\) will change with geometries, its sign will not, and the only invariant of the metric and its first derivative are functions of \(r \, \text{and} \, \e ^{f}\theta_{+}\theta_{-} \), or equivalently \(g^{ab}\partial_{a}r\partial_{b}r=-\frac{1}{2}\e ^{f}\theta_{+}\theta_{-}\), which has an important physical and geometrical meaning: a sphere is said to be trapped (untrapped) if \(\theta_{+}\theta_{-} >0\)(\(\theta_{+}\theta_{-}<0\)) and if \(\theta_{+}\theta_{-}=0\), it is a marginal sphere.
	
	Considering non-stationary black holes, Hayward proposed that the future outer trapping horizon is defined as the closure of a hypersurface foliated by a future or past, outer or inner marginal sphere, as a definition of a black hole, because this horizon has various properties that are typically intuitively attributed to the black hole, including observer limitations and analogues of the zeroth, first, and second laws of thermodynamics. However, in the case of the FRW universe, one should define the future trapping horizon as
	
	\begin{eqnarray} 
	\theta_{+}=0, \quad \theta_{-}<0, \quad \partial _{-}\theta _{+}>0,
	\label{2.3}
	\end{eqnarray}
	
	as a system for establishing thermodynamics, the surface gravity at the cosmological horizon is negative.
	
	In spherically symmetric spacetime, the total energy inside a sphere with radius \(r\) can be obtained by calculating the Minsner-Sharp energy:
	
	\begin{eqnarray} 
	E=\frac{r}{2G}\left (1-g^{ab}\partial _{a}r\partial _{b}r\right )=\frac{r}{G}\left ( \frac{1}{2} -g^{+-}\partial _{+}r\partial  _{-}r\right ),
	\label{2.4}
	\end{eqnarray}
	
	which is a pure geometric quantity that has much better properties than other energy definitions when considering non-stationary spacetime. The relation between the Minner-Sharp energy and other energies can be found in the Ref. \cite{Hayward_1938} . There are also two invariants constructed by the energy-momentum tensor \(T^{\mu \nu}\):
	
	\begin{eqnarray} 
	W=-\frac{1}{2}g_{ab}T^{ab}=-g_{+-}T^{+-},
	\label{2.5}
	\end{eqnarray}
	
	which is called the work density. \(\Psi\) is called the energy flux vector (also known as the energy-supply vector), and its components are
	
	\begin{eqnarray} 
	\Psi_{a}=T^{b}_{a}\partial _{b}r+W\partial _{a}r,
	\label{2.6}
	\end{eqnarray}
	
	Here and in the following, \(a,b\) represent the two-dimensional space normal to the sphere. According to the definition of Minsner-Sharp energy and the above two quantities, the (0,0) component of Einstein equations can be written as a unified first law:
	
	\begin{eqnarray} 
	dE=A\Psi + WdV,
	\label{2.7}
	\end{eqnarray}
	
	where \(A=4\pi r^{2}\) and\(V=\frac{4\pi}{3}r^{3}\). This unified law contains rich information, for example, by projecting the unified first law along the trapping horizon, we can obtain the first law of black hole thermodynamics, which takes the form \cite{AKBAR20067} 
	
	\begin{eqnarray} 
	\left <dE,\mathit{z}  \right >=\frac{\kappa}{8\pi G}\left <  dA,\mathit{z} \right>+\left < WdV,\mathit{z} \right>,
	\label{2.8}
	\end{eqnarray}
	
	where \(\kappa\) defined by
	
	\begin{eqnarray} 
	\kappa=\frac{1}{2}\nabla ^{a}\nabla_{a}r,
	\label{2.9}
	\end{eqnarray}
	
	is the surface gravity of the trapping horizon. Here, \(\mathit{z}=\mathit{z}^{+}\partial _{+}+\mathit{z}^{-}\partial _{-}\) is the vector tangent to the captured horizon. It should be noted that according to the definition of horizon \(\partial_{+}r=0\), we have
	
	\begin{eqnarray} 
	\mathit{z}^{a}\partial _{a}\left( \partial _{+}r \right)=\mathit{z}^{+}\partial _{+}\partial _{+}r+\mathit{z}^{-}\partial _{-}\partial _{+}r=0,
	\label{2.10}
	\end{eqnarray}
	
	on the rapping horizon, then
	
	\begin{eqnarray} 
	\frac{\mathit{z}^{-}}{\mathit{z}^{+}}=-\frac{\partial _{+}\partial _{+}r}{\partial _{-}\partial _{+}r},
	\label{2.11}
	\end{eqnarray}
	
	Also note that by using the Einstein equations \(\partial _{+}\partial _{+}r=-4\pi r T_{++}\)  \cite{Hayward_1938}, and the definition of the surface gravity in Eq. (\ref{2.9}). It is easily to find
	
	\begin{eqnarray} 
	\left< A\Psi ,\mathit{z} \right >=\frac{\kappa}{8\pi G}\left< dA , \mathit{z} \right >,
	\label{2.12}
	\end{eqnarray}
	
	which is the Clausius relation in the thermodynamic version of black holes, as shown in the first term on the right side of the Eq. (\ref{2.8}) . If the temperature is defined as \(T=\kappa /2\pi\) and the entropy is defined as \(S=A/4G\), then the left side of the above equation is only the heat flow \(\delta Q\), and the right side is the form \(TdS\). Therefore, in Einstein theory, the unified first law also contains the Clausius relation, which also holds in Gauss-Bonnet and Lovelock gravity theories by treating higher-order derivative terms as effective energy-momentum tensors. However, in the scalar-tensor theory, due to certain non-equilibrium thermodynamic properties, this relation no longer holds. In the next section, we will investigate the thermodynamics of the \(H^{2}+H^{-2}\) DE model and calculate the corresponding entropy.
	
	\section{THERMODYNAMICS OF THE \texorpdfstring{\(H^{2}+H^{-2}\)}{H\textasciicircum 2+H\textasciicircum -2} DARK ENERGY MODEL}\label{sec3}
	
	The research on Einstein equation is an effective way to investigate the dynamics of the universe. And we studied a (3+1)-dimensional flat Friedmann-Robertson-Walker (FRW) universe with a double-null form metric,
	
	\begin{eqnarray} 
	ds^{2}=h_{ab}dx^{a}dx^{b}+\tilde{r}^{2}d\Omega^{2},
	\label{3.1}
	\end{eqnarray}
	
	where \(x^{0}=t,x^{1}=r\), and \(\tilde{r}=a(t)r\), which is the radius of the sphere while \(a(t)\) is the scale factor. Defining
	
	\begin{eqnarray} 
	d\xi^{\pm}=-\frac{1}{\sqrt{2}}\left( dt\mp  \frac{a}{\sqrt{1-kr^{2}}}dr\right ),
	\label{3.2}
	\end{eqnarray}
	
	where \(k\) is the spacial curvature, and the metric could be rewritten as a double-null form
	
	\begin{eqnarray} 
	ds^{2}=-2d\xi^{+}d\xi^{-}+\tilde{r}^{2}d\Omega^{2},
	\label{3.3}
	\end{eqnarray}
	
	Let \(\partial _{+}\tilde{r}|_{\tilde{r}=\tilde{r}_{T}}=0\), then we will get the trapping horizon
	
	\begin{eqnarray} 
	\tilde{r}_{T}=\left(  H^{2}+\frac{k}{a^{2}} \right)^{-1/2},
	\label{3.4}
	\end{eqnarray}
	
	which has the same form as the apparent horizon, and \(H=\dot{a}/a\) is the Hubble parameter. Then we defined
	
	\begin{eqnarray} 
	\epsilon \equiv \frac{\dot{\tilde{r}}_{T}}{2H\tilde{r}_{T}},
	\label{3.5}
	\end{eqnarray}
	
	Therefore, we can obtain the ("inner") surface gravity
	
	\begin{eqnarray} 
	\kappa =-(1-\epsilon)/\tilde{r}_{T}=-\frac{1}{\tilde{r}_{T}}\left (  1-\frac{\dot{\tilde{r}}_{T}}{2H\tilde{r}_{T}} \right )=-\frac{\tilde{r}_{T}}{2}\left (  \dot{H} +2H^{2}+\frac{k}{a^{2}} \right ),
	\label{3.6}
	\end{eqnarray}
	
	Note that the surface gravity here is different from the surface gravity on the "outer" trapping horizon commonly defined. We assume that \( \epsilon < 1 \) and \( \kappa < 0 \) here, because we define an surface gravity on the  "inner" trapping horizon and define the corresponding temperature T (\( <0 \)).\cite{Andrei_2003,Danielsson_2005,Bousso_2005,Calcagni_2005,Cai_200503}
	
	In this case, it is obvious that \(\partial _{-}\tilde{r}_{T}<0\), therefore the trapping horizon is future.By using Eq. (\ref{2.10}) and directly calculating, when selecting \(\mathit{z}^{+}=1\), it can be obtained that \(\mathit{z}^{-}=\epsilon/(1-\epsilon)\). Then, in the \((t,r)\) coordinate system, the project vector is given by \(\mathit{z}=\partial_{t}-(1-2\epsilon)Hr\partial_{r}\).
	
	In the framework of \(H^{2}+H^{-2}\) DE \cite{Fang_2024}, add curvature and replace \(H^{2}\) with \(H^{2}+\frac{k}{a^{2}}\). For the convenience of calculation and processing, we introduce the following parameters
	
	\begin{eqnarray} 
	\alpha_{_G}=\frac{\alpha}{8 \pi G} \quad \tilde{\beta}_{_G}=\frac{\tilde{\beta}}{8 \pi G}=\frac{\beta H_{0}^{4}}{8 \pi G} \quad \rho_{m_{_G}}=8\pi G\rho_{m},
	\label{3.7}
	\end{eqnarray}
	
	Then, we can modify the Friedmann equation as
	
	\begin{eqnarray} 
	H^{2}+\frac{k}{a^{2}}=\frac{8\pi G}{3}\left [ \rho_{m}+3\alpha_{_G} \left( H^{2}+\frac{k}{a^{2}} \right)+3\tilde{\beta}_{_G} \left( H^{2}+\frac{k}{a^{2}} \right)^{-1} \right ],
	\label{3.8}
	\end{eqnarray}
	
	Solve this equation, we can obtain
	
	\begin{eqnarray} 
	H^{2}+\frac{k}{a^{2}}=\frac{8\pi G}{3}g\left(  \rho_{m} \right )=\frac{\rho _{m_{_G}}+\sqrt{\rho _{m_{_G}}^{2}+36\tilde{\beta}(1-\alpha)}}{6(1-\alpha)},
	\label{3.9}
	\end{eqnarray}
	
	where \(g\left(  \rho_{m} \right )\) is a function of the energy density of matter
	
	\begin{align} 
		\begin{split}
			g\left ( \rho_{m} \right )=\rho_{m}+\frac{3\alpha}{8 \pi G}\cdot \frac{8 \pi G \rho_{m}+\sqrt{(8 \pi G \rho_{m})^{2}+36\tilde{\beta}(1-\alpha)}}{6(1-\alpha)}\\+\frac{3\beta H_{0}^{4}}{8 \pi G}\cdot \frac{6(1-\alpha)}{8 \pi G \rho_{m}+\sqrt{(8 \pi G \rho_{m})^{2}+36\tilde{\beta}(1-\alpha)}},
			\label{3.10}
		\end{split}
	\end{align}
	
	Introduce a new reduced dimensionless parameter of energy density of matter
	
	\begin{eqnarray} 
	\tilde{\rho}_{m}=\frac{8\pi G}{6\sqrt{\tilde{\beta}(1-\alpha)}}\cdot\rho_{m}=\frac{8\pi G}{6H_{0}^{2}\sqrt{\beta(1-\alpha)}}\cdot\rho_{m}=\frac{8\pi G \rho_{m0}}{3H_{0}^{2}}\cdot\frac{(1+z)^{3}}{2\sqrt{\beta(1-\alpha)}}=\Omega_{m0}\frac{(1+z)^{3}}{2\sqrt{\beta(1-\alpha)}},
	\label{3.11}
	\end{eqnarray}
	
	Then, we can rewrite the Eq. (\ref{3.10}), and obtain the derivative of \(g\left(  \rho_{m} \right )\) with respect to \(\rho_{m}\)
	
	\begin{eqnarray} 
	g\left(  \rho_{m} \right )=\frac{3H_{0}^{2}}{8\pi G}\sqrt{\frac{\beta}{1-\alpha}}\left(  \tilde{\rho}_{m}+\sqrt{1+\tilde{\rho}_{m}^{2}} \right),
	\label{3.12}
	\end{eqnarray}
	
	\begin{eqnarray} 
	g'\left(  \rho_{m} \right )=\frac{1}{2(1-\alpha)}\cdot\left( 1+ \frac{\tilde{\rho}_{m}}{\sqrt{1+\tilde{\rho}_{m}^{2}}} \right) =\frac{1}{2(1-\alpha)}\cdot \frac{\tilde{\rho}_{m}+\sqrt{1+\tilde{\rho}_{m}^{2}}}{\sqrt{1+\tilde{\rho}_{m}^{2}}} ,
	\label{3.13}
	\end{eqnarray}
	
	Using the expression of \(g\left(  \rho_{m} \right )\) above, we can also express the relation between area and reduced dimensionless parameter of energy density of matter as
	
	\begin{eqnarray} 
	A=\frac{4\pi}{H^{2}+\frac{k}{a^{2}}}=\frac{4\pi}{\frac{8\pi G}{3}\cdot g(\rho_{m})}
	=\frac{4\pi}{H_{0}^{2}}\sqrt{\frac{1-\alpha}{\beta}}\frac{1}{\tilde{\rho}_{m}+\sqrt{1+\tilde{\rho}_{m}^{2}}},
	\label{3.14}
	\end{eqnarray}
	
	\begin{eqnarray} 
	dA=\frac{4\pi}{H_{0}^{2}}\sqrt{\frac{1-\alpha}{\beta}}\frac{\tilde{\rho}_{m}-\sqrt{1+\tilde{\rho}_{m}^{2}}}{\sqrt{1+\tilde{\rho}_{m}^{2}}},
	\label{3.15}
	\end{eqnarray}
	
	We defined the effect energy density \(\rho_{e}\), and using the continuity equations, we can obtain the effective pressure \(p_{e}\) corresponds to the effective energy density
	
	\begin{eqnarray} 
	\rho_{e}=g(\rho_{m})-\rho_{m}=\frac{3H_{0}^{2}}{8\pi G}\left [ (2\alpha -1)\frac{\beta^{1/2}}{(1-\alpha)^{1/2}}\tilde{\rho}_{m} +\sqrt{1+\tilde{\rho}_{m}^{2}} \right ],
	\label{3.16}
	\end{eqnarray}
	
	\begin{eqnarray} 
	p_{e}=(\rho_{m}+p_{m})g'-g-p_{m}=\frac{3H_{0}^{2}}{8\pi G}\frac{\beta^{1/2}}{(1-\alpha)^{1/2}}\left[   (\omega_{m}-2+2\alpha)\tilde{\rho}_{m}+\frac{\omega_{m}\tilde{\rho}_{m}^{2}-1}{\sqrt{1+\tilde{\rho}_{m}^{2}}}  \right ],
	\label{3.17}
	\end{eqnarray}
	
	Then, we obtain the relevant work density \(W_{e}\) and energy-supply vector \(\Psi_{e}\),
	
	\begin{align} 
		\begin{split} 
			W_{e}&=\frac{1}{2}\left [2g-\rho_{m}+p_{m}-(\rho_{m}+p_{m})g'\right ]\\
			&=\frac{3H_{0}^{2}}{8\pi G}\frac{\beta^{1/2}}{(1-\alpha)^{1/2}}\left [  \left ( \alpha - \frac{1}{2}\right )\left (1-\omega _{m}\right )\tilde{\rho}_{m} +\frac{1+\frac{1}{2}\left(  1-\omega _{m} \right) \tilde{\rho}_{m}^{2}}{\sqrt{1+\tilde{\rho}_{m}^{2}}} \right ],
			\label{3.18}
		\end{split}
	\end{align}
	
	\begin{align} 
		\begin{split} 
			\Psi_{e}&=\frac{1}{2}\left (g'-1\right )\left (\rho_{m}+p_m\right )\left( -H\tilde{r}_{T}dt+adr \right)\\
			&=\frac{3H_{0}^{2}}{8\pi G}\beta^{1/2}(1-\alpha)^{1/2}\left(  1+\omega _{m} \right)\left [ \frac{2\alpha-1}{2(1-\alpha)}+\frac{1}{2(1-\alpha)}\frac{\tilde{\rho}_{m}}{\sqrt{1+\tilde{\rho}_{m}^{2}} }\right ]\tilde{\rho}_{m}\left(  -H\tilde{r}_{T}dt+adr  \right ),
			\label{3.19}
		\end{split}
	\end{align}
	
	Identify \(T=\kappa /2\pi \), and use the relation
	
	\begin{eqnarray} 
	\dot{H}-\frac{k}{a^{2}}=-\frac{2\epsilon }{\tilde{r}_{T}^{2}}=-4\pi Gg'\left(  \rho_{m}+p_{m} \right ),
	\label{3.20}
	\end{eqnarray}
	
	we obtain
	
	\begin{eqnarray} 
	\delta Q_{e}=\left <  A\Psi_{e},\mathit{z} \right >=\frac{\kappa AH\epsilon }{2\pi G}\left (  \frac{g'-1}{g'} \right )=T\left (  \frac{g'-1}{4Gg'} \right )\left <  dA,\mathit{z} \right >,
	\label{3.21}
	\end{eqnarray}
	
	For the heat flow of pure matter, we also have
	
	\begin{eqnarray} 
	\delta Q_{m}=\frac{T}{4G g'}\left< dA,\mathit{z}   \right>=T\left<  dS_{m},\mathit{z}  \right >,
	\label{3.22}
	\end{eqnarray}
	
	where the entropy is obtained by
	
	\begin{align}
		\begin{split} 
			dS_{m}&=\frac{dA}{4Gg'}=\frac{dA}{4G}\cdot 2(1-\alpha)\frac{\sqrt{1+\tilde{\rho}_{m}^{2}}}{\tilde{\rho}_{m}+\sqrt{1+\tilde{\rho}_{m}^{2}}}\\
			&=\frac{2\pi}{GH_{0}^{2}}\frac{(1-\alpha)^{3/2}}{\beta^{1/2}}\frac{\tilde{\rho}_{m}-\sqrt{1+\tilde{\rho}_{m}^{2}}}{\tilde{\rho}_{m}+\sqrt{1+\tilde{\rho}_{m}^{2}}}d\tilde{\rho}_{m},
			\label{3.23}
		\end{split}
	\end{align}
	
	Compared to the usual entropy-area relation\((dS=dA/4G)\), there is a correction factor \( 2(1-\alpha)\frac{\sqrt{1+\tilde{\rho}_{m}^{2}}}{\tilde{\rho}_{m}+\sqrt{1+\tilde{\rho}_{m}^{2}}}\) in the entropy-area relation under this model.
	
	Therefore, we obtain the entropy by integrating the above Eq. (\ref{3.23})
	
	\begin{eqnarray} 
	S_{m}=\int \frac{dA}{4Gg'}=\frac{2\pi}{GH_{0}^{2}}\frac{(1-\alpha)^{3/2}}{\beta^{1/2}}\left [   \frac{2}{3}\left(  1+\tilde{\rho}_{m}^{2} \right)^{3/2} -\frac{1}{3} \tilde{\rho}_{m} \left (  3+2\tilde{\rho}_{m}^{2}  \right )  \right ],
	\label{3.24}
	\end{eqnarray}
	
	From Eq. (\ref{3.14}), we can get
	
	\begin{eqnarray} 
	\frac{A}{4G}=\frac{\pi}{G H_{0}^{2}}\frac{(1-\alpha)^{1/2}}{\beta^{1/2}}\frac{1}{\tilde{\rho}_{m}+\sqrt{1+\tilde{\rho}_{m}^{2}}},
	\label{3.25}
	\end{eqnarray}
	
	Then, we can obtain a new area law satisfied by entropy
	
	\begin{eqnarray} 
	S_{m}=(1-\alpha)\frac{A}{4G}+\frac{G^{2}H_{0}^{4}\beta}{3\pi ^{2}}\left(   \frac{A}{4G} \right)^{3},
	\label{3.26}
	\end{eqnarray}
	
	Compared to the usual area law \(S=A/4G\), there is a correction factor \((1-\alpha)\), and an additional area cubic term \(\frac{G^{2}H_{0}^{4}\beta}{3\pi ^{2}}\left(   \frac{A}{4G} \right)^{3}\) is added.

	\section{OBSERVATIONAL CONSTRAINTS AND EVOLUTIONARY PROPERTIES\label{sec4}}

	In this section, we will use the best fitting values of the \(H^{2}+H^{-2}\) DE model parameters from Ref. \cite{Fang_2024} to study the thermodynamic evolution properties of this model. Using the natural unit system, let \(8\pi G=1\), then Eq. (\ref{3.26}) and Eq. (\ref{3.25}) can be rewritten as:
	
	\begin{eqnarray} 
	S_{m}=(1-\alpha)\frac{A}{4G}+\frac{H_{0}^{4}\beta}{192\pi ^{4}}\left(  \frac{A}{4G}  \right)^{3},
	\label{4.1}
	\end{eqnarray}
	
	\begin{eqnarray} 
	\frac{A}{4G}=\frac{8 \pi ^{2}}{H_{0}^{2}}\frac{(1-\alpha)^{1/2}}{\beta^{1/2}}\frac{1}{\tilde{\rho}_{m}+\sqrt{1+\tilde{\rho}_{m}^{2}}},
	\label{4.2}
	\end{eqnarray}
	
	From\cite{Fang_2024}, we can get the best fitting values of the \(H^{2}+H^{-2}\) DE model parameters.
	
	\begin{eqnarray} 
	H_{0}=72.8 \quad \alpha =0.088 \quad \beta =0.686 \quad \Omega_{m0}=0.226,
	\label{4.3}
	\end{eqnarray}
	
	Substitute the above best fitting parameter values into Eq. (\ref{4.1}) and Eq. (\ref{4.2}), we can obtain
	
	\begin{eqnarray} 
	S_{m}=0.031332 \cdot \left[  \frac{2}{3}\left( 1+\tilde{\rho}_{m}^{2}  \right)^{3/2} -\frac{1}{3}\tilde{\rho}_{m}\left (  3+2\tilde{\rho}_{m}^{2} \right ) \right ],
	\label{4.4}
	\end{eqnarray}
	
	\begin{eqnarray} 
	\frac{A}{4G}=0.017178 \cdot \frac{1}{\tilde{\rho}_{m}+\sqrt{1+\tilde{\rho}_{m}^{2}}},
	\label{4.5}
	\end{eqnarray}
	
	\begin{eqnarray} 
	S_{m}=0.912\cdot \frac{A}{4G}+1030.26\cdot\left(  \frac{A}{4G} \right)^{3},
	\label{4.6}
	\end{eqnarray}
	
\begin{figure}[H]
		\centering
		\includegraphics[width=0.6\linewidth]{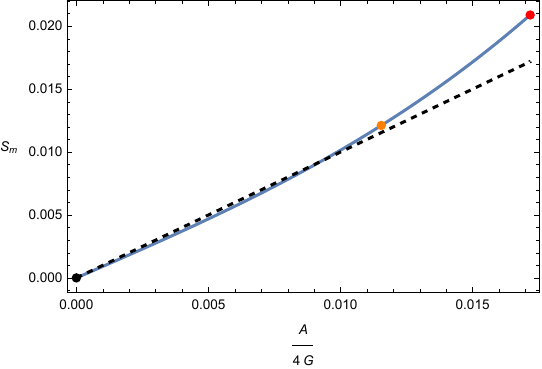}
		\caption{The entropy's variation with surface area is depicted, with the red point, orange point, and black point representing the initial universe, the energy density of matter equals that of dark energy, and the infinite future, respectively. \\
			The dashed line represents the typical entropy-area relationship.\\
			The parameters in the models are set to their best-fit values. }
		\label{ewithA}
	\end{figure}

\begin{figure}[H]
		\centering
		\includegraphics[width=0.4\linewidth]{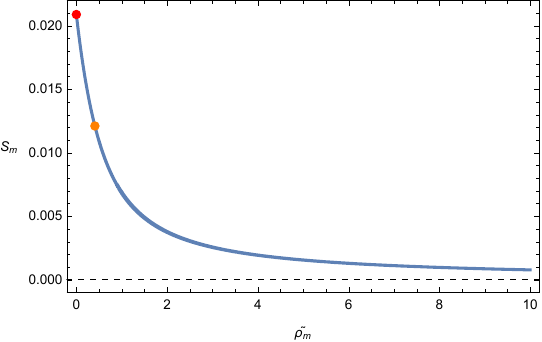}
		\includegraphics[width=0.4\linewidth]{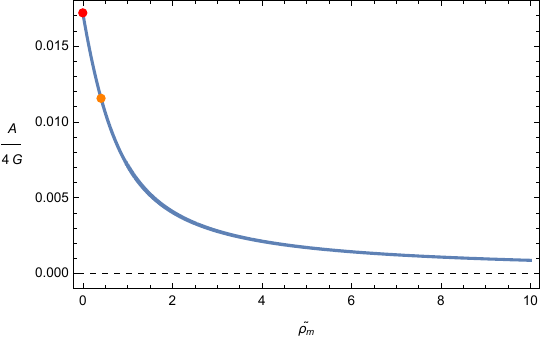}
		\caption{Left: The entropy's variation with the reduced dimensionless parameter of energy density of matter is depicted, with the orange point and red point representing the energy density of matter equals that of dark energy and the infinite future, respectively.And the dashed line marks the initial entropy of the universe. \\
			Right:The surface area's variation with the reduced dimensionless parameter of energy density of matter is depicted, with the orange point and red point representing the energy density of matter equals that of dark energy and the infinite future, respectively.And the dashed line marks the initial surface area of the universe.\\
			The parameters in the models are set to their best-fit values.}
		\label{eandA}
	\end{figure}

	Fig. \ref{ewithA} and Fig. \ref{eandA} provide evolutionary graphs between entropy \(S_{m}\), surface area \(A\), and the reduced dimensionless parameter of energy density of matter \(\tilde{\rho}_{m}\).
	
	We can see that in the early universe, when the reduced dimensionless parameter of energy density of matter \(\tilde{\rho}_{m}\) is large and the area \(A/4G\) is small, the law of entropy-area is approximately linear. However, as the universe evolves,  \(\tilde{\rho}_{m}\) gradually decreases and \(A/4G\) gradually increases, the entropy area law gradually deviates from linearity, which is a prominent feature caused by \(H^{2}+H^{-2}\) DE. When the universe reaches the end of its evolution, the area tends towards a finite maximum value \((A/4G)_{max}=0.017178\), while the entropy tends towards a finite maximum value \(\left (S_{m}\right )_{max}=0.020888\).
	
	We can also calculate the derivative of entropy with respect to time from Eq. (\ref{3.23}) and Eq. (\ref{3.11})
	
	\begin{eqnarray}
	\dot{S}_{m}=\left\{\begin{matrix}
	&-\frac{6\pi}{GH_{0}}\frac{(1-\alpha)^{\frac{5}{4}}}{\beta^{\frac{1}{4}}}\tilde{\rho}_{m}\cdot \frac{\tilde{\rho}_{m}-\sqrt{1+\tilde{\rho}_{m}^{2}}}{\left (\tilde{\rho}_{m}+\sqrt{1+\tilde{\rho}_{m}^{2}}   \right )^{\frac{1}{2}}}, k=0\\
	&-\frac{6\pi}{GH_{0}}\frac{(1-\alpha)^{\frac{5}{4}}}{\beta^{\frac{1}{4}}}\tilde{\rho}_{m}\cdot \frac{\tilde{\rho}_{m}-\sqrt{1+\tilde{\rho}_{m}^{2}}}{\tilde{\rho}_{m}+\sqrt{1+\tilde{\rho}_{m}^{2}}} \cdot \left [  \left(  \tilde{\rho}_{m}+\sqrt{1+\tilde{\rho}_{m}^{2}} \right) - \frac{1}{H_0^2}  \frac{(1-\alpha)^{\frac{5}{6}}}{\beta ^{\frac{1}{6}}} \cdot \left (   \frac{2}{\Omega_{m0}} \right ) ^{\frac{2}{3}} \cdot  \tilde{\rho}_{m}^{\frac{2}{3}} \right]^{\frac{1}{2}},k=+1\\
	&-\frac{6\pi}{GH_{0}}\frac{(1-\alpha)^{\frac{5}{4}}}{\beta^{\frac{1}{4}}}\tilde{\rho}_{m}\cdot \frac{\tilde{\rho}_{m}-\sqrt{1+\tilde{\rho}_{m}^{2}}}{\tilde{\rho}_{m}+\sqrt{1+\tilde{\rho}_{m}^{2}}} \cdot \left [  \left(  \tilde{\rho}_{m}+\sqrt{1+\tilde{\rho}_{m}^{2}} \right) + \frac{1}{H_0^2}  \frac{(1-\alpha)^{\frac{5}{6}}}{\beta ^{\frac{1}{6}}} \cdot \left (   \frac{2}{\Omega_{m0}} \right ) ^{\frac{2}{3}} \cdot  \tilde{\rho}_{m}^{\frac{2}{3}} \right]^{\frac{1}{2}},k=-1
	\end{matrix}\right.
	\label{4.7}
	\end{eqnarray}
	
	Similarly, in the natural unit system, use the best fitting parameter values Eq. (\ref{4.3}), these equations can be rewritten as
	
	\begin{eqnarray}
	\dot{S}_{m}=\left\{\begin{matrix}
	& -6.37268\cdot \tilde{\rho}_{m}\cdot
	\frac{\tilde{\rho}_{m}-\sqrt{1+\tilde{\rho}_{m}^{2}}}{\left (
	\tilde{\rho}_{m}+\sqrt{1+\tilde{\rho}_{m}^{2}}    \right )^{\frac{1}{2}}}, k=0\\
	& -6.37268\cdot \tilde{\rho}_{m}\cdot
	\frac{\tilde{\rho}_{m}-\sqrt{1+\tilde{\rho}_{m}^{2}}}{\tilde{\rho}_{m}+\sqrt{1+\tilde{\rho}_{m}^{2}}} \cdot \left ( \tilde{\rho}_{m}+\sqrt{1+\tilde{\rho}_{m}^{2}} -0.000796 \tilde{\rho}_{m}^{\frac{2}{3}} \right ) ^{\frac{1}{2}} , k=+1\\
	& -6.37268\cdot \tilde{\rho}_{m}\cdot
	\frac{\tilde{\rho}_{m}-\sqrt{1+\tilde{\rho}_{m}^{2}}}{\tilde{\rho}_{m}+\sqrt{1+\tilde{\rho}_{m}^{2}}} \cdot \left ( \tilde{\rho}_{m}+\sqrt{1+\tilde{\rho}_{m}^{2}} +0.000796 \tilde{\rho}_{m}^{\frac{2}{3}} \right ) ^{\frac{1}{2}} , k=-1
	\end{matrix}\right.
	\label{4.8}
	\end{eqnarray}
	It is obviously that the curvature parameter of the universe has a very small impact on the results.
	Meanwhile, by using Eq. (\ref{3.4}), (\ref{3.5}), (\ref{3.6}) ,and \(T=\kappa /2\pi\), and employing the best fitting parameter value Eq. (\ref{4.3}), the expression for temperature can be obtained as
	
	\begin{eqnarray}
	T=-10.79031\cdot\left (  \tilde{\rho}_{m}+\sqrt{1+\tilde{\rho}_{m}^{2}} \right )^{\frac{1}{2}}\cdot \left(  1-\frac{3}{4}\frac{\tilde{\rho}_{m}}{\sqrt{1+\tilde{\rho}_{m}^{2}}} \right),
	\label{4.9}
	\end{eqnarray}
	
\begin{figure}[H]
		\centering
		\includegraphics[width=0.6\linewidth]{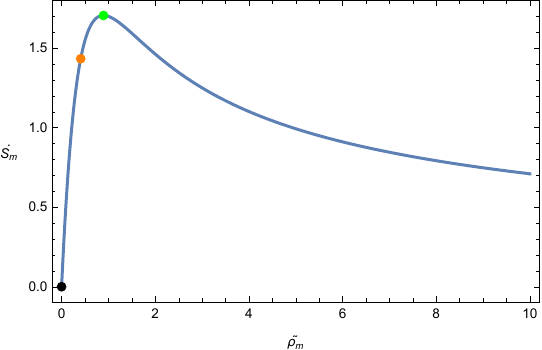}
		\caption{The variation of derivative of entropy with respect to time with the reduced dimensionless parameter of energy density of matter is depicted, with the black point, orange point, and green point representing the infinite future, the energy density of matter equals that of dark energy, and the minimum derivative of entropy, respectively. \\
			The parameters in the models are set to their best-fit values. }
		\label{edot}
	\end{figure}

\begin{figure}[H]
		\centering
		\includegraphics[width=0.6\linewidth]{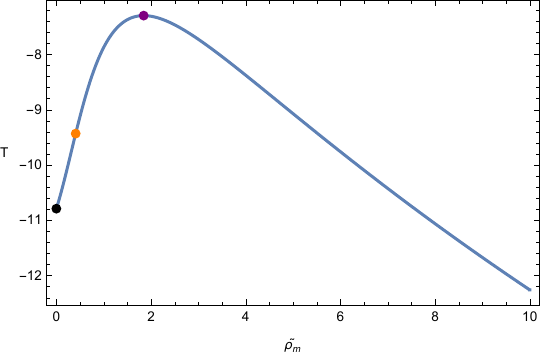}
		\caption{The variation of temperature with the reduced dimensionless parameter of energy density of matter is depicted, with the black point, orange point, and purple point representing the infinite future,the energy density of matter equals that of dark energy, and the minimum temperature, respectively. \\
			The parameters in the models are set to their best-fit values. }
		\label{T}
	\end{figure}

	Fig. \ref{edot} and Fig. \ref{T} provide evolutionary graphs between the derivative of entropy with respect to time \(\dot{S}_{m}\), temperature \(T\), and the reduced dimensionless parameter of energy density of matter \(\tilde{\rho}_{m}\).
	
	We can see that the derivative of entropy with respect to time \(\dot{S}_{m}\) first undergoes a slow growth process, reaches the maximum value \(\left(\dot{S}_{m}\right)_{max}=1.70467\) at \(\tilde{\rho}_{m}=\frac{2\sqrt{5}}{5}\) , \(z=0.84308\), and then asymptotically decreases to 0. And the temperature \(T\) reaches the extreme value \(T_{max}=-7.29886\) when \(\tilde{\rho}_{m}=\sqrt{\frac{25+3\sqrt{57}}{14}}\), \(z=1.34613\), and then asymptotically becomes a fixed value \(T=-10.79031\).
	
	For now \(z=0\), \(\tilde{\rho}_{m}=\Omega_{m0}\cdot\frac{1}{2\sqrt{(1-\alpha)\beta}}=0.14286\), therefore the entropy change rate and temperature are respectively: \(\dot{S}_{m}= 0.73534 \quad \text{and} \quad T=-10.3575\).

	\section{CONCLUSIONS AND DISCUSSIONS}
	
	In this paper, we investigated the thermodynamic properties of the \(H^{2}+H^{-2}\) DE model. This is a DE model based on the holographic principle, inspired by the first-order approximation of Kaniadakis entropy, and using the Hubble horizon \(1/H\) as the infrared cutoff. This DE model can explain the accelerated expansion of the universe, solve the Hubble tension problem, and avoid the "big rip" problem. In this stage of research, we derived the entropy-area relation of this DE model 
	
	\[dS_{m}=\frac{dA}{4G}\cdot 2(1-\alpha)\frac{\sqrt{1+\tilde{\rho}_{m}^{2}}}{\tilde{\rho}_{m}+\sqrt{1+\tilde{\rho}_{m}^{2}}}\]
	
	and compared with the usual entropy-area relation, there is a correction factor \(2(1-\alpha)\frac{\sqrt{1+\tilde{\rho}_{m}^{2}}}{\tilde{\rho}_{m}+\sqrt{1+\tilde{\rho}_{m}^{2}}}\). We also obtained the revised area law 
	
	\[S_{m}=(1-\alpha)\frac{A}{4G}+\frac{G^{2}H_{0}^{4}\beta}{3\pi ^{2}}\left(   \frac{A}{4G} \right)^{3}\]
	
	If \(\alpha = 0\) and \(\beta =0 \), then the above equation returns to the usual law of area \(S = \frac{A}{4G}\). And compared to the usual area law, this revised area law has both a correction factor \((1-\alpha)\) and an additional area cubic term \(\frac{G^{2}H_{0}^{4}\beta}{3\pi ^{2}}\left(   \frac{A}{4G} \right)^{3}\).
	
	We also conducted a quantitative analysis of the thermodynamic evolution of this model using the best fitting parameter values obtained from our previous studies Ref. \cite{Fang_2024}. Through analysis, it was found that the area cannot grow infinitely, but there exists an upper limit \((A/4G)_{max}=0.017178\); In the early universe, the entropy-area relation was approximately linear, gradually deviating and eventually tending towards a fixed maximum \(\left(  S_{m} \right)=0.020888\); The time rate of change of entropy is always positive and reaches its maximum value \(\left(\dot{S}_{m}\right)_{max}=1.70467\) when \(z=0.84308\); In addition, the temperature also reaches the extreme value \(T_{max}=-7.29886\) at \(z=1.34613\).
	
	This paper starts from the thermodynamics of the universe and provides a possible origin of \(H^{2}+H^{-2}\) DE. In this theory, the universe contains only limited information. However, we still haven't traced this \(H^{2}+H^{-2}\) DE model back to a specific, mathematically modified theory of gravity, which will be our next research goal.

	\section{Acknowledgments}
	This work is supported by National Science Foundation of China grant No. 11105091.
	

	\bibliographystyle{unsrt}
	\bibliography{refs}
	
\end{document}